\documentclass[%
 reprint,
superscriptaddress,
 amsmath,amssymb,
 aps,
]{revtex4-2}

\usepackage{graphicx}
\usepackage{dcolumn}
\usepackage{bm}
\usepackage{hyperref}
\hypersetup{colorlinks,linkcolor={blue},citecolor={blue},urlcolor={black}}

\begin{document}

\preprint{APS/123-QED}

\title{Two-dimensional THz spectroscopy of nonlinear phononics \\ in the topological insulator MnBi\textsubscript{2}Te\textsubscript{4} }
\author{T.G.H. Blank}
\affiliation{Radboud University, Institute for Molecules and Materials, 6525 AJ Nijmegen, The Netherlands.}
\author{K.A. Grishunin}
\affiliation{Radboud University, Institute for Molecules and Materials, 6525 AJ Nijmegen, The Netherlands.}
\author{K.A. Zvezdin}
\affiliation{Prokhorov General Physics Institute of the Russian Academy of Sciences, 119991 Moscow, Russia.}
\author{N.T. Hai}
\affiliation{Department of Physics, National Changhua University of Education, Changhua 500, Taiwan.}
\author{J.C. Wu}
\affiliation{Department of Physics, National Changhua University of Education, Changhua 500, Taiwan.}
\author{S.-H. Su}
\affiliation{Department of Physics, National Cheng Kung University, Tainan 701, Taiwan.}
\author{J.-C. A. Huang}
\affiliation{Department of Physics, National Cheng Kung University, Tainan 701, Taiwan.}
\author{A.K. Zvezdin}
\affiliation{Prokhorov General Physics Institute of the Russian Academy of Sciences, 119991 Moscow, Russia.}
\author{A.V. Kimel}
\affiliation{Radboud University, Institute for Molecules and Materials, 6525 AJ Nijmegen, The Netherlands.}

\date{\today}
\begin{abstract}
The interaction of a single-cycle THz electric field with the topological insulator MnBi\textsubscript{2}Te\textsubscript{4} triggers strongly anharmonic lattice dynamics, promoting fully coherent energy transfer between the otherwise non-interacting Raman-active $E_g$ and infrared (IR)-active $E_u$ phononic modes. Two-dimensional (2D) THz spectroscopy combined with modeling based on the classical equations of motion and symmetry analysis reveals the multi-stage process underlying the excitation of the Raman-active $E_g$ phonon. In this process, the THz electric field first prepares a coherent IR-active $E_u$ phononic state and subsequently interacts with this state to efficiently excite the $E_g$ phonon. 
\end{abstract}

\maketitle

Although the lattice dynamics of real crystals are rather complex and imply multiple mutual correlations between movements of individual atoms, the modern theory of condensed matter has successfully managed to describe the dynamics in terms of linear superpositions of mutually independent phononic modes. However, if the amplitude of the lattice vibrations is large, this approximation fails and the lattice dynamics enter a poorly explored regime of nonlinear phononics \cite{Forst2011, nicoletti2016nonlinear, Disa2021}. In this regime, a single phonon can be excited by multiple photons and electromagnetic excitation opens up new channels of energy transfer between otherwise non-interacting phononic modes \cite{PhysRevB.2.4294, PhysRevB.3.2063, https://doi.org/10.1002/pssb.2220610214, PhysRevB.35.9278, Forst2015}. Thanks to the recent developments of intense single-cycle THz pulses \cite{Hebling:02, Hoffmann_2011, hirori2011}, the electric field of the latter can exceed $1$~MV/cm and thus has become comparable with the interatomic fields ($\sim 100$~MV/cm) \cite{kittel1996introduction, PhysRevB.80.165203}. While such exceptionally intense pulses must facilitate strongly nonlinear interaction of the THz electromagnetic fields with the lattice \cite{SALEN20191, PhysRevB.89.220301, PhysRevX.11.021067}, there are very few experimental reports on the nonlinear excitation of phonons in this spectral range \cite{Forst2011, PhysRevB.97.174302, PhysRevLett.122.073901, Kozina2019, Disa2020, PhysRevLett.129.207401}. 

It appears that the family of topological insulators originating from the parent compound Bi\textsubscript{2}Te\textsubscript{3} (or Bi\textsubscript{2}Se\textsubscript{3}) form an excellent playground to explore nonlinear phononics experimentally \cite{Nature1}. Due to the heavy bismuth and tellurium ions, these materials feature several phononic modes in the THz spectral range \cite{PhysRevB.5.3171, Rauh1981,C9CP01494B,  https://doi.org/10.1002/jrs.6255, Pei_2020, Choe2021, PhysRevB.103.L121103, cho2021phonon, ALIEV2019443, PhysRevB.105.214304, nano11123322, Padmanabhan2022}. For instance, it was previously demonstrated that Raman-active phonons in Bi\textsubscript{2}Se\textsubscript{3}, which cannot be excited by the electric field of light directly, may still be driven by THz electric field via a nonlinear mechanism \cite{PhysRevB.97.214304}. In particular, it was suggested that the nonlinear excitation occurs due to anharmonicity of the lattice dynamics of the IR-active phonons. Here, we employ two-dimensional (2D) THz spectroscopy to deduce information about the nonlinear THz-light-lattice interactions in MnBi\textsubscript{2}Te\textsubscript{4}. Using the experimental results and simulations we prove that the nonlinear excitation of a Raman-active phonon proceeds as a double-stage process. First, the THz electric field excites an IR-active phonon and prepares a coherent phononic state. Afterward, the THz electric field interacts with this coherent state to transfer energy from the IR-active to the Raman-active phonon.

We studied a $10$~nm film of ($0001$)-oriented MnBi\textsubscript{2}Te\textsubscript{4} grown by molecular beam epitaxy on a c-plane $\sim 300$~$\mu$m thick Al\textsubscript{2}O\textsubscript{3} substrate (see Ref.~\cite{nano11123322} for all the details of the growth procedure and characterization of the sample). The sample was capped by $4$~nm of MgO to prevent it from oxidizing. The material consists of Van der Waals (VdW) coupled septuple layers (SLs) of MnBi\textsubscript{2}Te\textsubscript{4} stacked along the out-of-plane direction ($z$-axis). We excited the sample with intense single-cycle THz pulses that were generated by titled pulse-front optical rectification in a LiNbO\textsubscript{3} crystal \cite{Hebling:02, hirori2011}, using near-infrared (NIR) pulses with a central wavelength of $800$~nm, $4$~mJ pulse energy, $1$~kHz repetition rate and $100$~fs pulse duration. Using off-axis parabolic mirrors, the generated THz pulses were guided and tightly focused onto the sample. The sample was mounted on a cold-finger cryostat where it can be cooled using liquid helium. The THz pulses were incident perpendicular to the sample plane, along the VdW stacking $z$-axis, such that the THz electric field has only in-plane components. The peak THz electric field of $780$~kV/cm was calibrated using electro-optical sampling in a $50$~$\mu$m ($110$)-oriented GaP crystal \cite{doi:10.1063/1.123601} (see Supplemental Material~S$1$~\footnote{\label{footnote}See Supplemental Material for details on the calibration of THz pulses in GaP, for the single THz pump induced waveforms as a function of temperature, pump and probe polarization, and THz amplitude, for details of the point-group $D_{3d}$ and for a detailed treatment of analytical and numerical modeling of the nonlinear excitation of the Raman-active phonon, which includes Refs. \cite{doi:10.1063/1.123601, 1591367, C9CP01494B, altmann1994point, PhysRevB.102.104102, PhysRevB.105.214304, PhysRevLett.122.073901, loudon, PhysRevLett.118.054101, https://doi.org/10.1002/jrs.6255, PhysRevX.11.021067}.}). A weak linearly polarized NIR probe pulse traveled collinear to the THz pump and was focused within the THz spot on the sample. The initial probe polarization was set orthogonal to the THz electric field. Using a balanced detection scheme and a data-acquisition card \cite{doi:10.1063/1.3669783}, we tracked the THz pump-induced rotation of the probe polarization $\theta(t)$ as a function of the time delay $t$ between the THz pump and probe pulse. 

\begin{figure}[t!]
    \centering
    \includegraphics{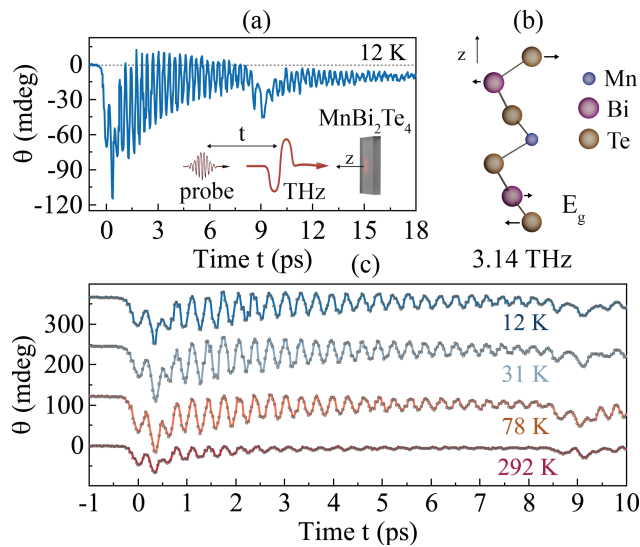}
    \caption{\small{(a) Dynamics induced by a single THz pulse 
that is measured by changes in the polarization rotation of the probe, performed at a temperature of $T = 12$~K. We attribute the second dip at $t\approx 9$~ps to a THz reflection in the substrate. The frequency of the mode corresponds to that of an $E_g$ Raman-active phonon. (b) Single SL of the VdW stacked MnBi\textsubscript{2}Te\textsubscript{4} material, showing the displacements of the ions for the Raman-active $E_g$ phonon mode \cite{Padmanabhan2022}. (c) Waveforms recorded at various temperatures ranging from $12$~K to room temperature, showing marginal differences.}} 
    \label{fig:1}
\end{figure}

A typical polarization rotation transient induced by a single THz pulse at a temperature of $12$~K is shown in Fig.~\ref{fig:1}(a). Two apparent features stand out: oscillation at a frequency of $3.14$~THz ($104.7$~cm\textsuperscript{-1}) and a THz-induced offset with respect to the unperturbed state. Also, we observe a second spike in the signal after $t\approx 9$~ps, which we attribute to the round-trip reflection of the initial THz pulse in the substrate. The frequency of the observed mode corresponds to that of a double degenerate Raman-active $E_g$ mode, previously measured in MnBi\textsubscript{2}Te\textsubscript{4} \cite{Pei_2020, Choe2021, cho2021phonon, ALIEV2019443, PhysRevB.105.214304, nano11123322} or its parent compound Bi\textsubscript{2}Te\textsubscript{3} \cite{C9CP01494B} by Raman spectroscopy. It corresponds to in-plane vibrations of the bismuth and tellurium ions as depicted schematically in Fig.~\ref{fig:1}(b) \cite{Padmanabhan2022}. We excited and detected highly similar waveforms in a broad range of temperatures from $12$~K to room temperature (see Fig.~\ref{fig:1}(c)). The observed temperature dependence of the oscillation frequency (see Supplemental Material Fig.~S$2$ \cite{Note1}) is in agreement with what is expected for the $E_g$ phonon \cite{C9CP01494B}.

Note that the spectrum of the THz pulse is sufficiently broad to potentially excite the infrared-active (IR-active) $E_u$ phonon at a frequency of $\sim 1.5$~THz \cite{PhysRevB.105.214304, PhysRevB.103.L121103}. However, only Raman-active modes of the $E_g$ species are able to induce birefringence in the medium, which can be explained via the mechanism of so-called ''depolarized scattering'' \cite{PhysRevLett.122.073901, PhysRev.147.608} (see Supplemental Material S$3$ \cite{Note1}). The resulting time-dependent birefringence induces ellipticity of the probe light, which is subsequently converted to a detectable rotation as a result of static linear birefringence in the sample and substrate. Therefore, although IR-active $E_u$ modes could be excited directly by our THz pulses, the experimental setup is only sensitive to Raman-active $E_g$ phonons.

\begin{figure}[b!]
    \centering
    \includegraphics{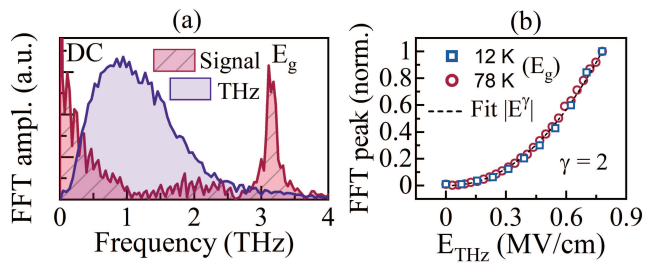}
    \caption{\small{(a) Fourier amplitudes of the signal (Fig.~\ref{fig:1}(a)) measured at $T = 12$~K (red), plotted together with the excitation spectrum of the single-cycle THz pulse (purple). (b) Normalized peak Fourier amplitude of $E_g$ mode as a function of the applied THz electric field, measured at $T = 12$~K and $78$~K. The fitting emphasizes that the amplitude of the $E_g$ mode scales quadratically with the exciting THz field amplitude.}}
    \label{fig:2}
\end{figure}

It is remarkable that the frequency of the observed $E_g$ Raman-active phonon lies completely beyond the spectrum of the exciting THz pulse (see Fig. \ref{fig:2}(a)), which implies that the excitation mechanism is nonlinear. The nonlinearity of the excitation can be further confirmed by measuring the peak Fourier amplitude of the mode at several powers of the THz electric field $E_\mathrm{THz}$, and afterward fitting this dependence by the power-law $|E_\mathrm{THz}|^\gamma$ for variable $\gamma$. The results are shown in Fig \ref{fig:2}(b) for the temperatures of $12$ and $78$~K. It shows that the amplitude of the phonon mode scales quadratically $\gamma = 2$ with the applied THz field. 

\begin{figure*}[t!]
    \centering
    \includegraphics{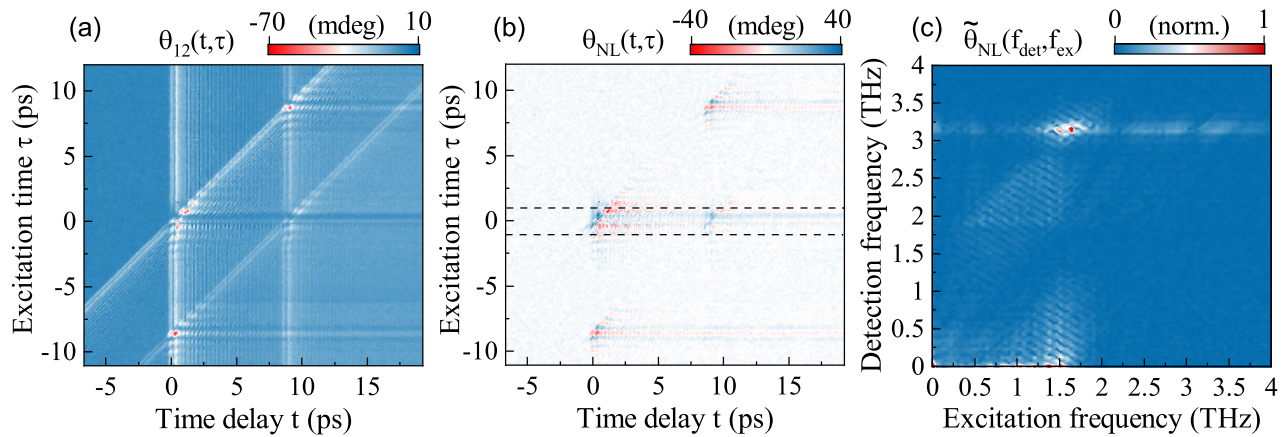}
    \caption{\small{(a) 2D mapping of the polarization rotation $\theta_{12}(t,\tau)$ as a function of $t$ and pump-pump delay $\tau$. Along the initial overlap and the modulations due to the $E_g$ mode, one can clearly see the internally reflected pulses as we observed in Fig. \ref{fig:1}. (b) The extracted nonlinear contribution $\theta_{\mathrm{NL}}(t, \tau)$ to the signal. The largest degree of nonlinearity is observed when the pulses (reflected or initial) overlap. In further analysis, we excluded the region between $\tau = \pm 1$ (indicated by the black dashed lines) to avoid artifacts as a result of the simultaneous presence of two pulses in the THz generating LiNbO\textsubscript{3} crystal (see Supplemental Material S$4$ \cite{Note1} for a detailed explanation). (c) 2D FFT of the non-linear signal $\theta_{\mathrm{NL}}(t,\tau)$. The result reveals two peaks at the detection frequency of the $E_g$ phonon ($3.14$~THz), at the excitation frequencies of approximately $1.47$ and $1.67$~THz. The former frequency can be assigned to that of an IR-active $E_u$ phonon, while the latter is at the complementing frequency $f_{E\textsubscript{g}} - f_{E\textsubscript{u}} $.}}
    \label{fig:3}
\end{figure*}

Theoretically, as the $E_g$ phonon is Raman-active and MnBi\textsubscript{2}Te\textsubscript{4} has a centrosymmetric crystal structure, the excitation of the phonon by a THz electric field is by symmetry only allowed via a nonlinear mechanism. The interaction between phonons and light is usually described by an interaction potential $U$, expressed in terms of the electric field of light and of the normal coordinates $Q_\mathrm{R}$ for Raman-active phonons and $Q_{\mathrm{IR}}$ for IR-active phonons. In linear approximation, we may only include ordinary electric dipole interaction (or electric linear polarizability) of the IR-active phonons, while the interaction energy of the electric field with the Raman-active phonons is strictly equal to zero.  

To account for the excitation of a coherent Raman-active phonon, we need to include terms beyond the linear approximation \cite{PhysRevX.11.021067}. In the simplest case, the nonlinearity in the thermodynamic potential can be represented by a term that accounts for the high amplitudes of the THz pump electric field $E$. This term is proportional to $Q_\mathrm{R}E^2$ and results in a driving force $\mathrm{d}U/\mathrm{d}Q_\mathrm{R} \propto E^2$ of the Raman-active mode \cite{PhysRevLett.119.127402}. In the case of ultrashort THz pulses, this force will act as a ''photonic impact''.
Other types of nonlinearity originate from the anharmonic response of the lattice to a THz electric field \cite{Forst2011, PhysRevLett.118.054101, PhysRevB.89.220301, PhysRevX.11.021067}. Suppose that the ultrashort THz pulse resonantly excites an IR-active mode $Q_{\mathrm{IR}}$ by the electric dipole interaction described by the interaction term $Z^* Q_{\mathrm{IR}}E$, where $Z^*$ is the mode effective charge. When such an IR-phonon is present, two additional nonlinear terms can be added to the thermodynamic potential, which are proportional to $Q_{\mathrm{R}}Q_{\mathrm{IR}}E$ and $Q_\mathrm{R}Q_{\mathrm{IR}}^2$. The term $\propto Q_\mathrm{R}Q_{\mathrm{IR}}E$ captures how a Raman-active phonon affects the polarization induced by IR-active phonons, and enables a nonlinear excitation of the Raman-active mode by THz-mediated coupling to the IR-active mode $Q_\mathrm{IR}$ \cite{PhysRevX.11.021067}. In other words, the THz electric field first resonantly drives an IR-active mode and subsequently interacts with this mode to promote energy transfer from the IR-active to the Raman-active mode. The other nonlinear interaction term, proportional to $Q_\mathrm{R}Q_{\mathrm{IR}}^2$, results from the natural anharmonicity of the interatomic potential, and couples Raman-active phonons to IR-active phonons without the need for an external electric field \cite{Forst2011, PhysRevLett.118.054101, PhysRevB.89.220301}. The driving force of the Raman-active mode in this case is proportional to the square of the amplitude of the IR-active mode $\mathrm{d}U/\mathrm{d}Q_\mathrm{R} \propto Q_{\mathrm{IR}}^2$. Altogether, when considering one IR-active and one Raman-active mode in a centrosymmetric medium, the interaction potential beyond linear approximation would thus acquire the following form:
\begin{equation}
    U = - Z^* Q_{\mathrm{IR}} E - bQ_\mathrm{R} Q_{\mathrm{IR}} E  - c Q_{\mathrm{R}}Q_{\mathrm{IR}}^2 - \delta_R Q_R E^2.
\end{equation}
For the Raman and IR-active phonons ($\nu = \mathrm{R}, \mathrm{IR}$), where $b, c$ and $\delta_R$ are phenomenological constants. The corresponding equations of motion are of the form:
\begin{equation}
\label{eq:Raman1}
\ddot Q_{\nu} +  \Gamma_\nu  \dot Q_{\nu} + \omega_{\nu}^2  Q_{\nu} =  -\frac{\delta U}{\delta Q_\nu},
\end{equation}
where $\Gamma_\nu = 2\zeta_\nu \omega_\nu$ are the damping coefficients and $\zeta_\nu$ are the dimensionless damping ratios.

In order to reveal the origin of the experimentally observed nonlinear excitation of the $E_g$ phonon, we performed two-dimensional (2D) THz spectroscopy \cite{Woerner_2013, Lu2018, PhysRevLett.122.073901}. This technique involves a pump-probe measurement using two THz pump pulses separated by a time $\tau$ referred to as \textit{excitation time}. The rotation $\theta_{12}(t,\tau)$ induced by the two THz pulses was measured as a function of both probe-delay $t$ and the excitation time $\tau$. We kept one of the two THz pulses fixed and delayed the other THz pulse for a time $\tau$ with respect to the fixed pulse. The two-dimensional mapping of this measurement is shown in Fig. \ref{fig:3}(a). It clearly shows 
the oscillations which we assigned to the $E_g$ mode, as well as the signals associated with the reflected THz pulses. At the same time, our experimental setup tracked the signals from pump $1$ ($\theta_1(t,\tau)$) and pump $2$ ($\theta_2(t,\tau))$ separately. Combining the three signals allowed us to calculate the - by definition - nonlinear contribution to the signal \cite{Lu2018, doi:10.1063/5.0046664, doi:10.1063/5.0047700, mashkovich2021terahertz}:
\begin{equation}
    \label{eq:INL}
    \theta_{\mathrm{NL}}(t,\tau) = \theta_{12}(t,\tau) - \theta_1(t,\tau) - \theta_2(t,\tau).
\end{equation} 
The result of this measurement is shown in Fig. \ref{fig:3}(b), emphasizing the large degree of nonlinearity of the $E_g$ mode. 

Finally, we analyzed this result in more detail by taking the two-dimensional fast Fourier transform (2D FFT) of $\theta_{\mathrm{NL}}(t,\tau)$, to obtain the normalized 2D Fourier amplitude $\widetilde{\theta}_{\mathrm{NL}}(f_{\mathrm{det}}, f_{\mathrm{ex}})$, where the detection frequency $f_{\mathrm{det}}$ and the excitation frequency $f_{\mathrm{ex}}$ are the conjugate variables of $t$ and $\tau$, respectively. The result for $\widetilde{\theta}_{\mathrm{NL}}(f_{\mathrm{det}}, f_{\mathrm{ex}})$ is shown in Fig.~\ref{fig:3}(c). We observe two peaks at the detection frequency $f_{\mathrm{det}} = 3.14$~THz of the $E_g$ mode, for the excitation frequencies of approximately $f_{\mathrm{ex}} = 1.47$~THz and $1.67$~THz. These peaks show that the excitation efficiency of the $E_g$ mode is modulated as a function of $\tau$. The frequencies of the modulation disclose the processes responsible for the nonlinear excitation of the $E_g$ mode. We note that the sum of these two frequencies equals that of the $E_g$ mode ($3.14$~THz). Moreover, we can assign the peak at $1.47$~THz to an $E_u$ IR-active phonon mode as was calculated by DFT \cite{PhysRevB.105.214304} and measured by IR-spectroscopy \cite{PhysRevB.103.L121103}. Hence the experimental results clearly show that the nonlinear excitation of the Raman-active $E_g$ phonon is practically mediated by the excitation of the IR-active $E_u$ phonon. 

\begin{figure}[t!]
    \centering
    \includegraphics{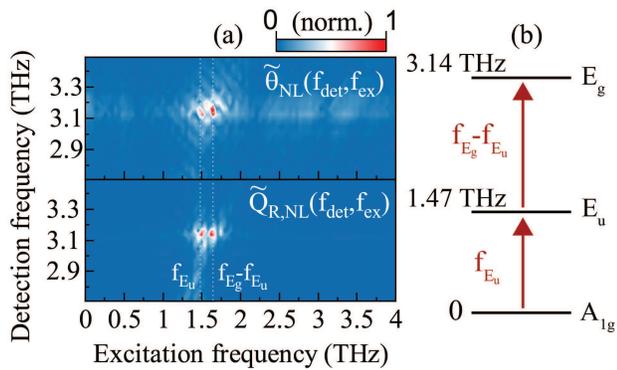}
    \caption{\small{(a) 2D FFTs of the experimental 
signal $\theta_\mathrm{NL}(t,\tau)$ and of the nonlinear part of the simulated $Q_\mathrm{R, NL}(t,\tau)$ modeled by the equations of motion \eqref{eq:Raman1}, taking the linear interaction term $\sim Z^* Q_{\mathrm{IR}}E$ and only the nonlinear interaction term $b Q_\mathrm{R} Q_\mathrm{IR}E$ into account. The result confirms that the excitation of the Raman-active $E_g$ phonon is mediated by the IR-active $E_u$ phonon. (b) Schematic illustration of the excitation mechanism in a model with two oscillators, corresponding to the $E_u$ phonon and $E_g$ phonon. The arrows indicate the stimulated transitions. }}
    \label{fig:4}
\end{figure}

To compare the theory of phonon dynamics with the experiment, we simulated the two coupled equations of motion (Eq.~\ref{eq:Raman1}) using realistic shapes of the THz pulses modeled by the Gaussian derivative function (see Supplemental Material S$5$ \cite{Note1}). In Fig.~\ref{fig:4}(a) we compare the experimental 2D FFT $\tilde{\theta}_{\mathrm{NL}}(f_{\mathrm{det}}, f_{\mathrm{ex}})$ with the simulated one $Q_\mathrm{R}(f_{\mathrm{det}}, f_{\mathrm{ex}})$, when taking only the nonlinear coupling term $bQ_\mathrm{R} Q_{\mathrm{IR}} E$ into account. It shows the high degree of similarity to the experiment that can be obtained in this case. The results of the modeling are supported by analytical solutions of the equations of motion (see Supplemental Material S$5$ \cite{Note1})). Based on the comparison of theory and experiment, we can definitively conclude that the dominant pathway of excitation of the $E_g$ Raman-active phonon relies on the generation of a coherent $E_u$ phonon by the THz electric field, and the subsequent interaction of the latter with this coherent $E_u$ phonon. 

In summary, we showed that a single-cycle THz electric field triggers strongly anharmonic lattice dynamics, which initiates an interaction between otherwise non-interacting phonons. The discovery of these strong lattice anharmonicities in the topological antiferromagnet MnBi\textsubscript{2}Te\textsubscript{4} in the THz spectral range opens up a plethora of opportunities for the field of nonlinear phononics and the lattice engineering of topological matter \cite{Disa2021}, enabling new routes for the ultrafast manipulation of complex quantum phases.

\begin{acknowledgments}
The authors acknowledge E.A. Mashkovich for his contribution to the development of the $2$D THz spectroscopy experimental setup. The authors thank S. Semin and Ch. Berkhout for technical support. The work was supported by de Nederlandse Organisatie voor Wetenschappelijk Onderzoek (NWO). 
\end{acknowledgments}


\bibliography{references}

\end{document}


\title{Supplemental material: Two-dimensional THz spectroscopy of nonlinear phononics in the topological insulator MnBi\textsubscript{2}Te\textsubscript{4} }
\author{T.G.H. Blank}
\affiliation{Radboud University, Institute for Molecules and Materials, 6525 AJ Nijmegen, The Netherlands.}
\author{K.A. Grishunin}
\affiliation{Radboud University, Institute for Molecules and Materials, 6525 AJ Nijmegen, The Netherlands.}
\author{K.A. Zvezdin}
\affiliation{Prokhorov General Physics Institute of the Russian Academy of Sciences, 119991 Moscow, Russia.}
\author{N.T. Hai}
\affiliation{Department of Physics, National Changhua University of Education, Changhua 500, Taiwan.}
\author{J.C. Wu}
\affiliation{Department of Physics, National Changhua University of Education, Changhua 500, Taiwan.}
\author{S.-H. Su}
\affiliation{Department of Physics, National Cheng Kung University, Tainan 701, Taiwan.}
\author{J.-C. A. Huang}
\affiliation{Department of Physics, National Cheng Kung University, Tainan 701, Taiwan.}
\author{A.K. Zvezdin}
\affiliation{Prokhorov General Physics Institute of the Russian Academy of Sciences, 119991 Moscow, Russia.}
\author{A.V. Kimel}
\affiliation{Radboud University, Institute for Molecules and Materials, 6525 AJ Nijmegen, The Netherlands.}

\date{\small{\today}}

\maketitle

\tableofcontents

\newpage

\section{THz electric field calibration by electro-optic sampling in GaP}
\begin{figure}[h!]
    \centering
    \includegraphics{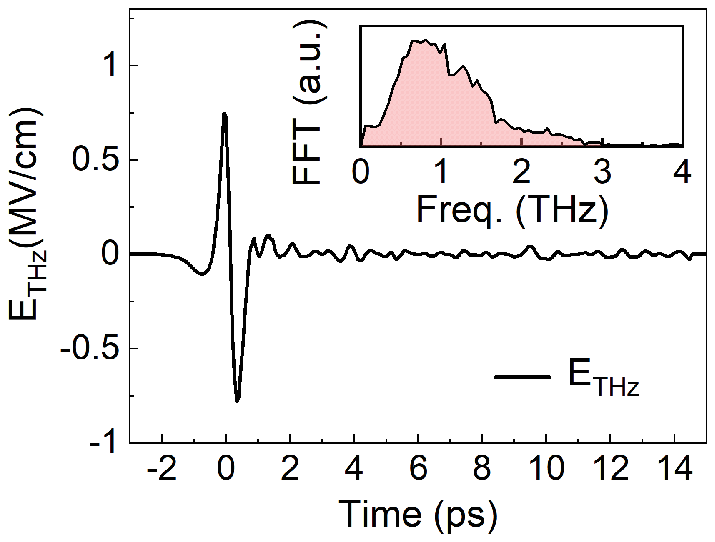}
    \caption{\textit{\small{THz calibration using electro-optical sampling in GaP. The inset shows the associated THz spectrum.}}}
    \label{fig:THz_callib}
\end{figure}
After the generation using tilted pulse front optical rectification in a LiNbO\textsubscript{3} crystal, the THz beam was expanded, collimated, and focused using three off-axis parabolic mirrors onto a ($110$) GaP crystal of $50$~$\mu$m thickness that was grown on a ($100$)-cut GaP substrate with $2$ mm thickness. The THz electric field induces birefringence in the ($110$) GaP \cite{doi:10.1063/1.123601, 1591367}, which is mapped as a function of time by measuring the changes of ellipticity of a linearly polarized probe pulse a central wavelength of $800$ nm in a balanced detection scheme. From the electro-optic coefficients, we calculated the THz electric field strength as in Fig. \ref{fig:THz_callib} resulting in a peak electric field strength of $780$ kV/cm.
\newpage

\section{Supplementary results of single THz pump-probe measurements}
The THz-induced dynamics shown in main the article have been characterized as a function of temperature $T$, the strength of the THz electric field, and the polarization of the pump and probe. Some of these dependencies are already shown in the main article. In this section, all the supplementary results of these dependencies are presented. 
\subsection{Temperature dependence}
\begin{figure}[ht!]
    \centering
    \includegraphics[scale = 1.0]{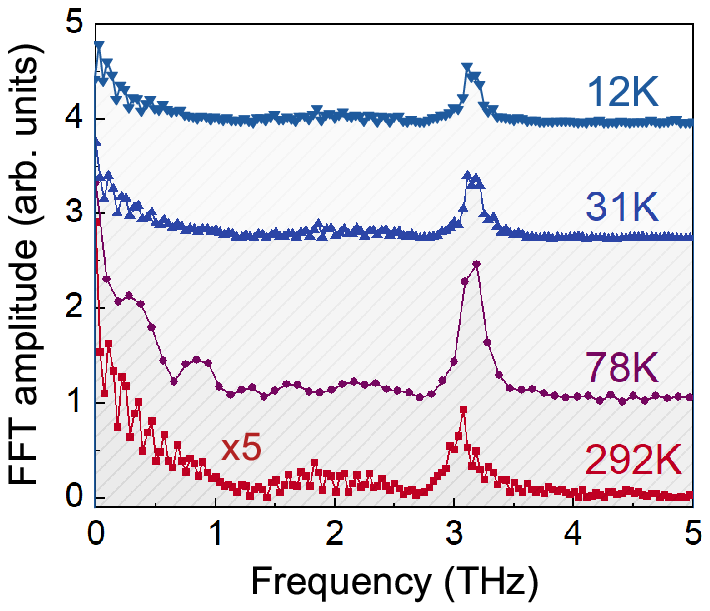}
\caption{\textit{\small{FFT spectrum of the THz-induced dynamics for several temperatures. The $E_g$ phonon mode frequency slightly decreases for increasing temperatures, in agreement with previous results \cite{C9CP01494B}. Moreover, the signal at $T = 78$~K is strong, making it suitable for the measurement using 2D THz spectroscopy. }} }
    \label{fig:Tscan}
\end{figure}

Figure \ref{fig:Tscan} shows the spectra of the waveforms for several temperatures, revealing a minor red-shift of the phonon frequency.
\newpage
\subsection{THz electric field strength dependence}
\begin{figure}[ht!]
    \centering
    \includegraphics[scale = 0.9]{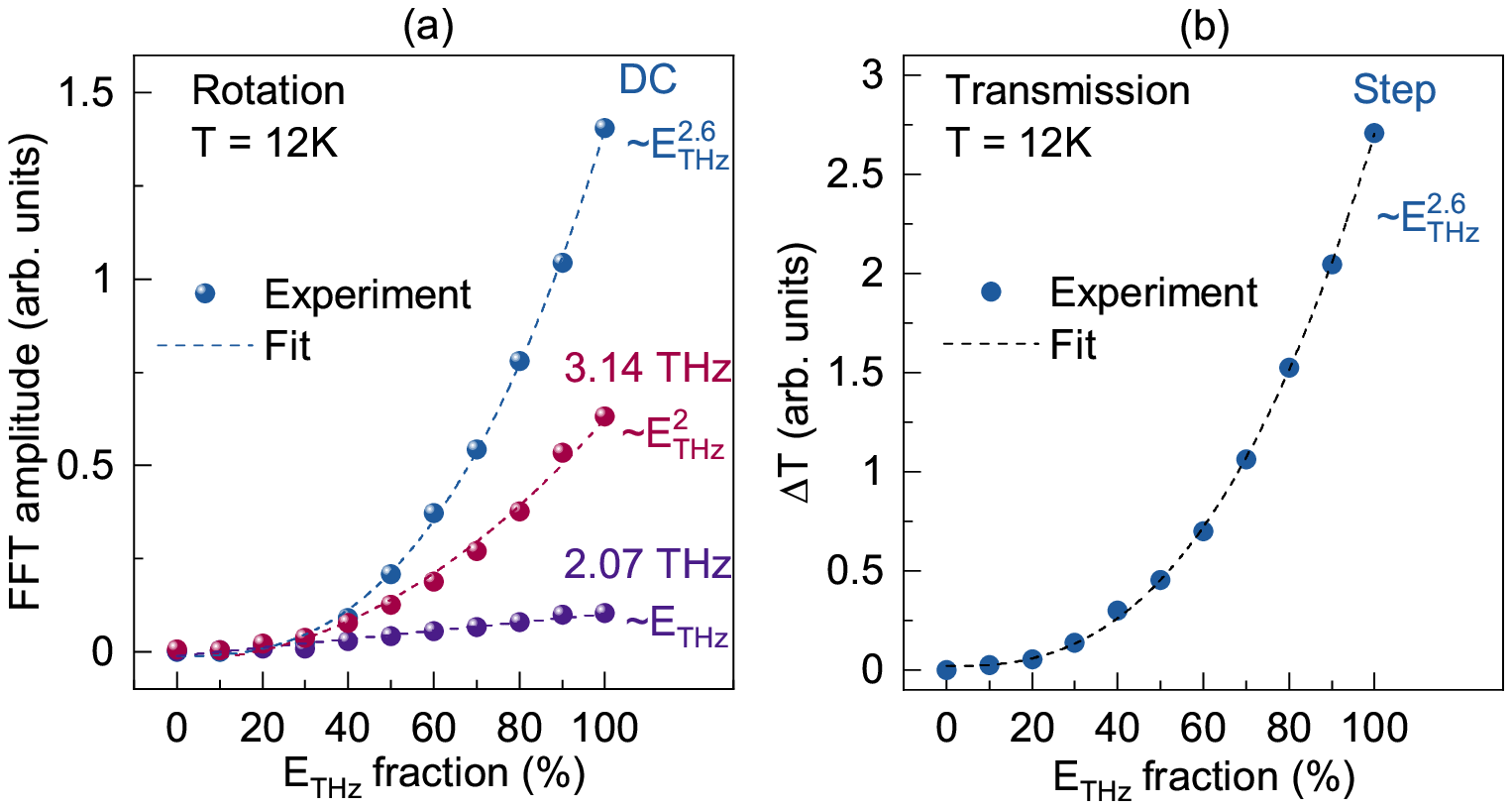}
    \caption{\textit{\small{(a) The FFT peak amplitude of the THz-induced rotation as a function of THz peak electric field strength (maximally $780$ kV/cm), measured at a temperature of $T = 12$~K. (b) Power-law dependence of the transient change of transmission, which is shown to have the same unusual power-law dependence $\gamma = 2.6$ as the DC component measured in the rotation data.}}}
    \label{fig:powerscan_12K}
\end{figure}

\begin{figure}[ht!]
    \centering
    \includegraphics[scale = 0.9]{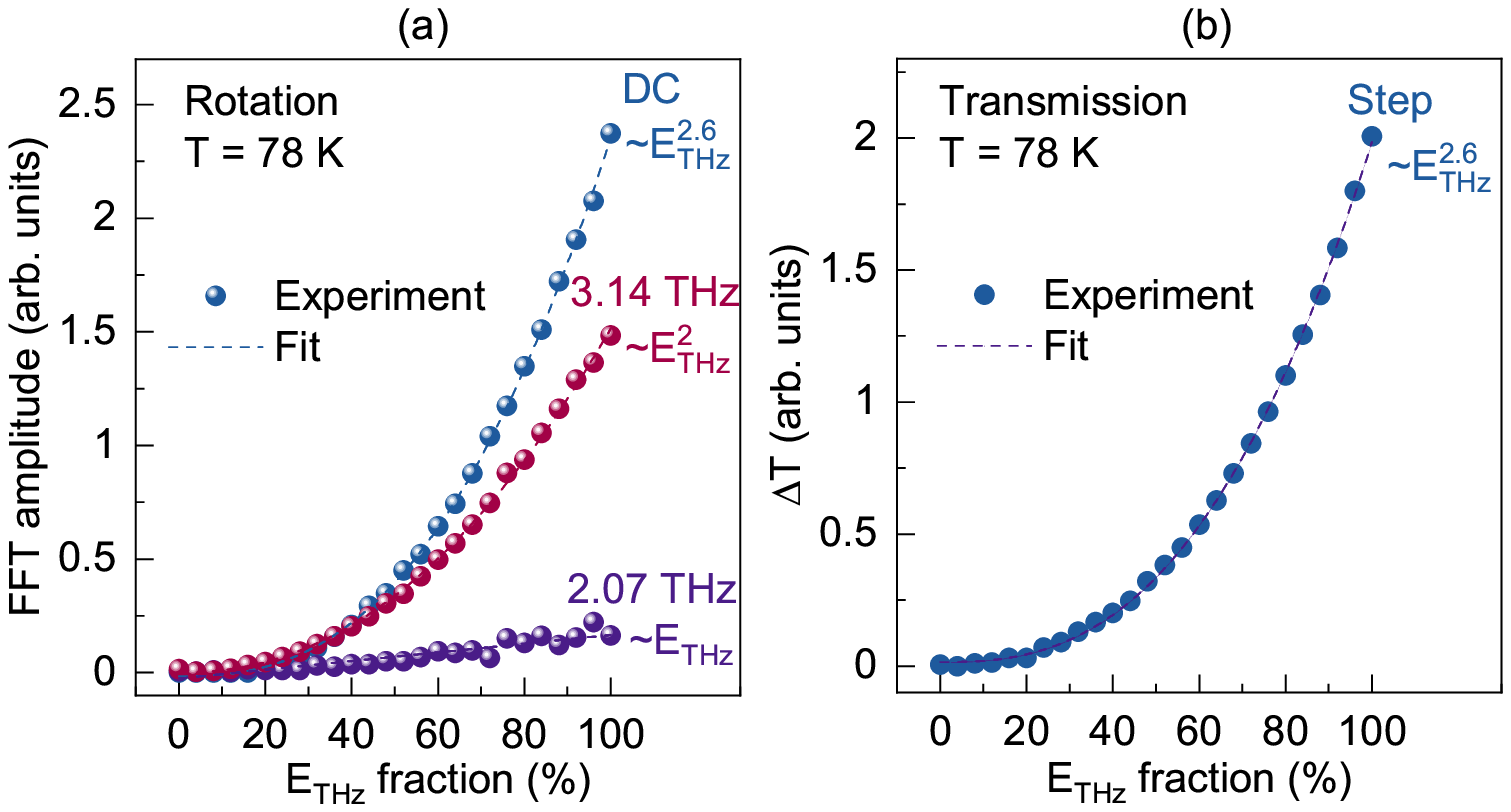}
    \caption{\textit{\small{Same measurement as described in Fig. \ref{fig:powerscan_12K}, but measured $T = 78$~K. The power-law dependencies w.r.t. $T= 12$~K are unchanged. }}}
    \label{fig:powerscan_78K}
\end{figure}

Most of the following results have already been plotted in the main article. However, we also noted a small contribution in the signal spectrum centered around $2.07$~THz, which we attribute to the THz-induced birefringence at the overlap (i.e. Pockels effect). Indeed, this contribution scaled exactly linear with the applied field (see Figs.~\ref{fig:powerscan_12K}(a) and \ref{fig:powerscan_78K}(a)). Secondly, we got a strong DC component in the spectrum due to THz-induced offset. This offset scales with a rather unusual power-law dependence $\propto E_\mathrm{THz}^{2.6}$.

Adding to this, besides measuring changes in the rotation of the probe, we also tracked changes in transmission. In all measurements, a step-like change in transmission was observed right when the THz pulse arrived. The relative change in transmission $\Delta T$ scales with exactly the same anomalous power-law dependence $\propto E_{\mathrm{THz}}^{2.6}$ as the THz-induced DC offset (see Figs. \ref{fig:powerscan_12K}(b) and \ref{fig:powerscan_78K}(b)), indicating a relation between the two.

\newpage
\subsection{Polarization dependencies of the THz pump ($\alpha$) and probe ($\beta$)}
\begin{figure}[ht!]   
\centering
\subfloat[$\beta = 0^\circ$.]{\includegraphics[width=0.3\textwidth, keepaspectratio]{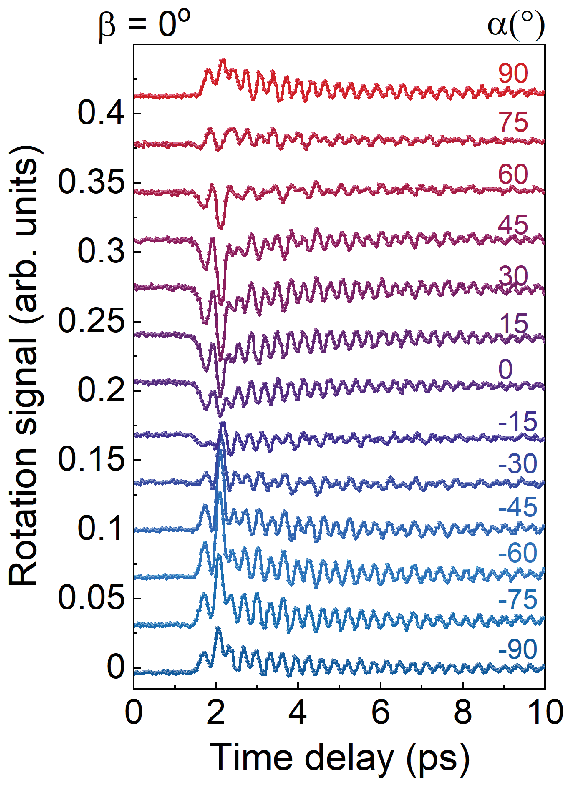}\label{fig: THz polar dep at 0probe}}
\subfloat[$\beta = 45^\circ$. ]{\includegraphics[width=0.3\textwidth, keepaspectratio]{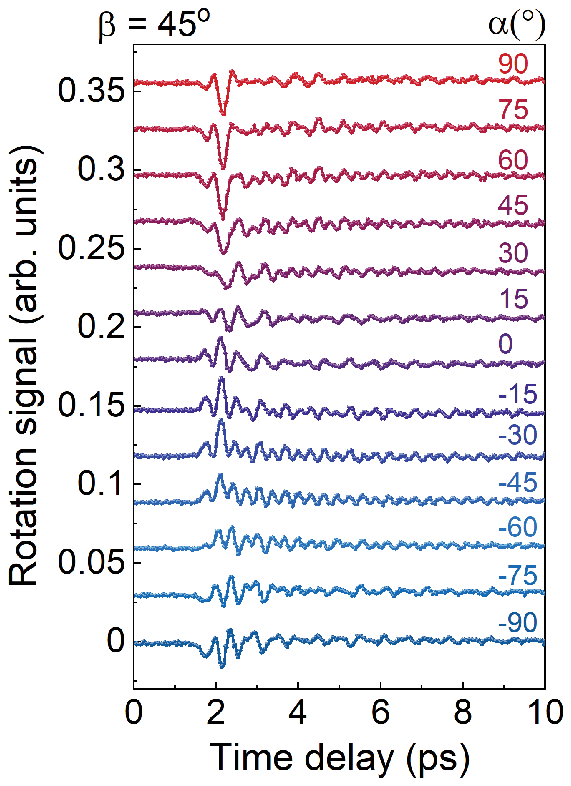}\label{fig: THz polar dep at 45probe}}
\subfloat[$\beta = 90^\circ$. ]{\includegraphics[width=0.3\textwidth, keepaspectratio]{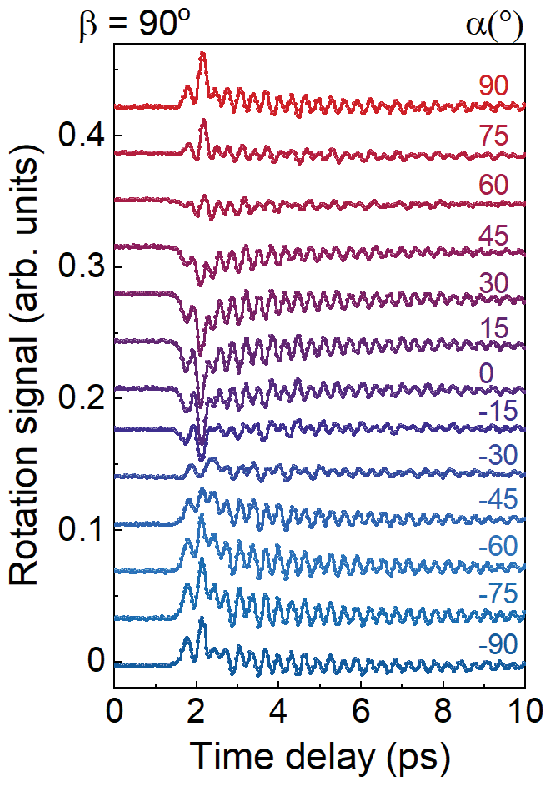}\label{fig: THz polar dep at 90probe}}
\caption[Optional caption for list of figures 5-8]{\textit{\small{Pump polarization scans for several probe polarizations $\beta = 0, 45 $ and $90$ degrees, measured at $T = 78$~K. The resemblance of (a) and (c) underlines a $\pi/2$ periodicity of the probe polarization, which will be confirmed in Fig.~\ref{fig:s7}. }  }}
\label{fig:Pumppol_severalprobepol}
\end{figure}

\begin{figure}[ht!]
    \centering
    \includegraphics{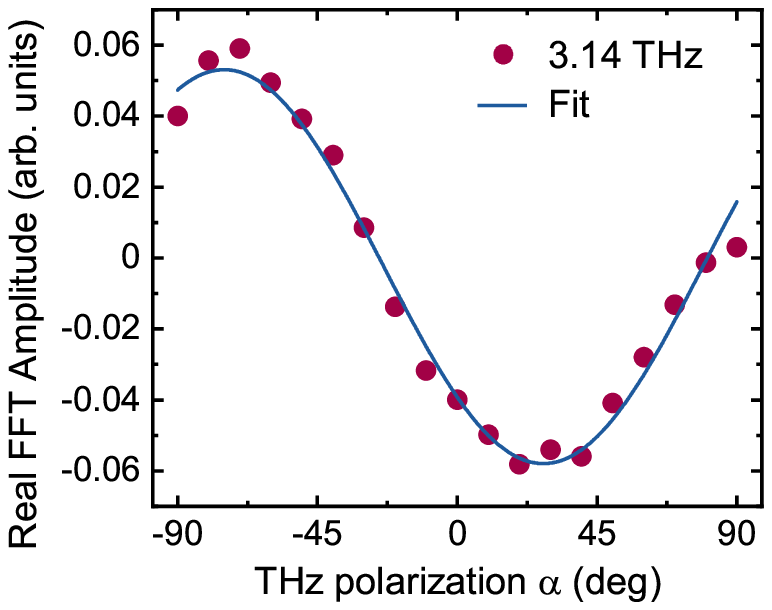}
    \caption{\textit{\small{Real FFT amplitudes of the data shown in Fig. \ref{fig:Pumppol_severalprobepol}(a), exhibiting the $\pi$ periodicity with respect to the pump polarization. It shows that the signal at $\alpha = 0^\circ$ is relatively strong, therefore it was used throughout in the main article to avoid rotating the THz polarization (which is accompanied by a loss in the field). }}}
    \label{fig:pumpscansT12K}
    \end{figure}

\begin{figure}[ht!]
    \centering
    \includegraphics{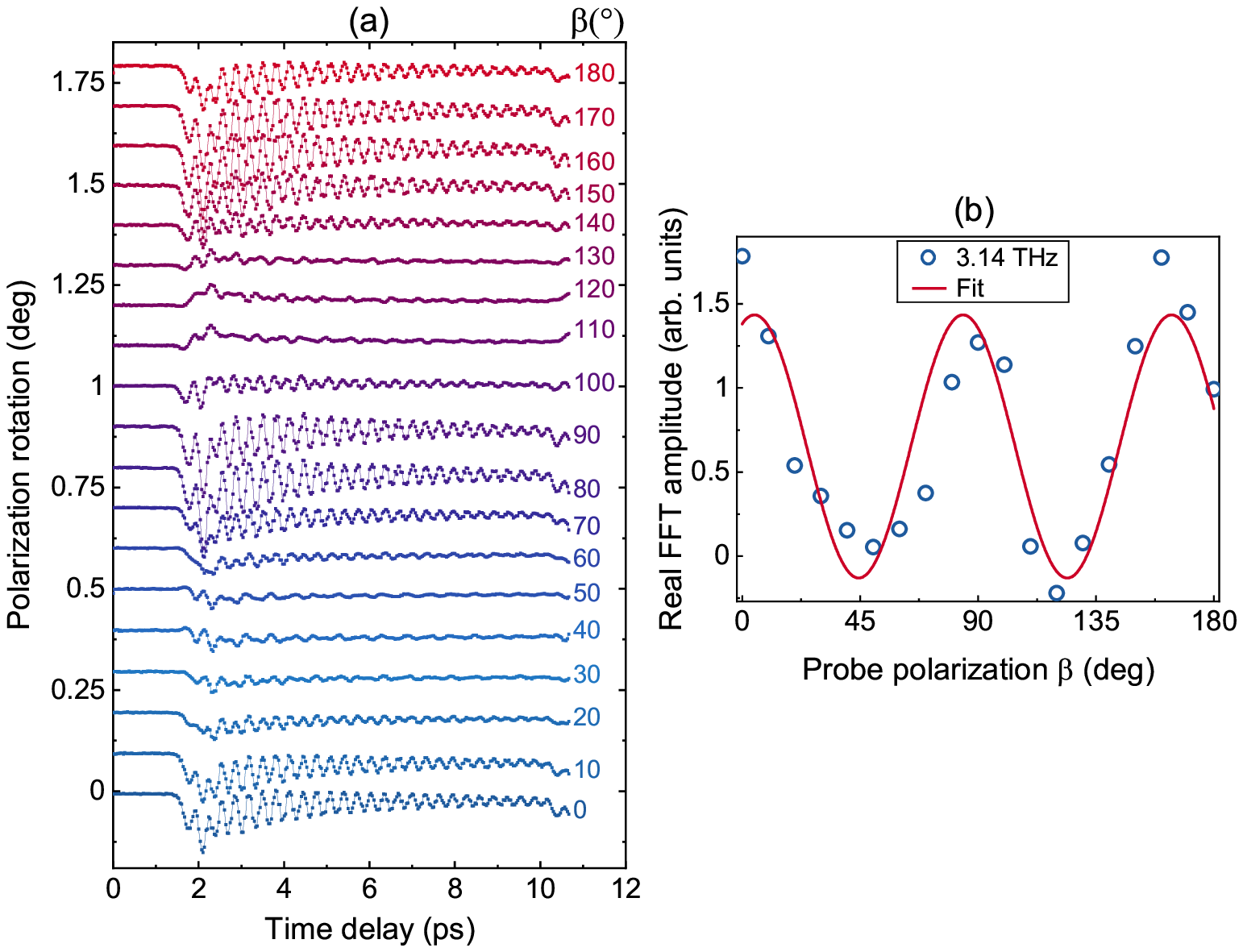}
    \caption[scale = 0.8]{\textit{\small(a) Different waveforms for several $\beta$ with $\alpha = 0^\circ$ fixed and $T = 78$~K. (b) Real FFT amplitude at $3.14$ THz as a function of probe polarization $\beta$. }}
    \label{fig:s7}
\end{figure}

The orientation of the THz pulse was controlled by a pair of wire-grid polarizers, and the probe polarization was rotated using a half-wave plate. The THz electric field orientation is characterized by the angle $\alpha$ with respect to the experimental $y$-axis $\mathbf{E}_{\mathrm{THz}} = E_0(\sin\alpha, \cos\alpha, 0)$, while the angle of the probe pulse electric field $\beta$ is defined with respect to the experimental $x$-axis $\mathbf{E}_{\mathrm{probe}} = E_{\mathrm{probe}}(\cos\alpha, \sin\alpha)$. We have no information about how these experimental axes compare to the crystal $x$ and $y$-axes. However, we do know that the pulses travel along the crystallographic $z$-axis, so the experimental $x-y$ plane does coincide with the crystallographic $x-y$ plane apart from an in-plane rotation. In all measurements presented in the main article, both $\alpha$ and $\beta$ were put to $0^\circ$, as these parameters gave strong results (see Figs.~\ref{fig:Pumppol_severalprobepol}, \ref{fig:pumpscansT12K} and \ref{fig:s7}) without needing to rotate the THz pulse which is accompanied by a loss in power as discussed below.

Rotating the THz polarization using two wire-grid polarizers reduces the peak THz electric field. In particular, the maximum reduction occurs when rotating to $\alpha = 90^\circ$, where the field is halved. Of course, during the scans as a function of THz polarization $\alpha$, we kept the field constant by picking the right angles of the wire-grid polarizers.

\newpage 

\newpage

\newpage

 \section{Group theory of the point group $D_{3d}$}
\begin{table}[h!]
 \caption{Character table for point-group $D_{3d}$ (rhombohedral) \cite{altmann1994point}.}
 \begin{center}
\begin{tabular}{c | cccccc|c|c }
    \hline
        & $E$ & $2C_3$ & $3C'_2$ & $i$ & $2iC_3$ & $3iC'_2$ & & \\
    \hline
    $A_{1g}$ & 1 & 1 & 1 & 1 & 1 & 1 & & $x^2 + y^2, z^2 $\\
    $A_{2g}$ & 1 & 1 & -1 & 1 & 1 & -1 & $R_z$ & \\ 
    $E_{g}$ & 2 & -1 & 0 & 2 & -1 & 0 & $(R_x, R_y)$ & $(xz, yz), (x^2 - y^2, xy)$ \\
    $A_{1u}$ & 1 & 1 & 1 & -1 & -1 & -1 & & \\
    $A_{2u}$ & 1 & 1 & -1 & -1 & -1 & 1 & $z$ &  \\
    $E_u$ & 2 & -1 & 0 & -2 & 1 & 0 & $(x, y)$ & \\
    \hline
\end{tabular}
\end{center}
\label{table_character}
\end{table}
From the basis functions in the character table associated with each irreducible representation, it can be seen that $A_{1g}$ and $E_g$ modes are Raman-active, while $A_{2u}$ and $E_u$ are IR-active. Given that MnBi\textsubscript{2}Te\textsubscript{4} has $7$ atoms in the unit cell, group theory predicts the existence of nine Raman-active $3A_{1g} \oplus 3E_g$ and nine IR-active optical phonons $3A_{2u} \oplus 3E_u$~\cite{PhysRevB.102.104102, PhysRevB.105.214304}.
\subsection{Detection mechanism}
Regarding the detection via the depolarized scattering mechanism, a Raman-active phonon may scatter light with a rotated polarization, if it transforms as any of the off-diagonal components of the polarizability tensor \cite{PhysRevLett.122.073901}. Given that the polarizability tensor is a symmetric tensor, these components must transform as $xy, xz$, and $yz$. Therefore, from the character table, it can be seen that only the $E_g$ mode contributes to the depolarized scattering mechanism. The scattering efficiency $S_{ij}$, and thereby the degree of ellipticity induced on the light, of a Raman-active mode with symmetry $\Delta$ can be expressed in terms of its Raman tensor \cite{loudon}:
\begin{equation}
    S_{ij} \propto ( \mathbf{e}_i \cdot \mathcal{R}(\Delta) \cdot \mathbf{e}_j)^2
\end{equation}
where $\mathcal{R}(\Delta)$ the Raman tensor of the $\Delta$ phonon and $\mathbf{e}_i$ and $\mathbf{e}_j$ are the unit polarization vectors of the incident and scattered light, respectively. For the two-fold degenerate $E_g$ mode there are two Raman tensors, and the total intensity can be found by adding the scattering efficiencies of them together \cite{loudon}:
\begin{equation}
    \mathcal{R}(E_g^{(I)}) = \begin{pmatrix}
        c & 0 & 0 \\
        0 & -c & d \\
        0& d & 0
    \end{pmatrix} \ \ \ \ \ \ \ \ \ \ \mathcal{R}(E_g^{(II)}) = \begin{pmatrix}
        0 & -c & -d \\
        -c & 0 & 0 \\
        -d& 0 & 0
    \end{pmatrix}
\end{equation}
where the $c,d $ are constants. On the contrary, the Raman tensor of $A_{1g}$ contains only diagonal components \cite{loudon}. 

However, in the experiment, we measured rotation and not ellipticity. We believe that the ellipticity induced by the depolarized scattering mechanisms is converted to a rotation due to static birefringence in the sample and substrate. 

\subsection{Symmetry-allowed coupling terms}

\begin{table}[b!]
\caption{Direct product table of the irreducible representations \cite{altmann1994point}.}
\begin{center}
\begin{tabular}{c|cccccc}
    \hline
        $\bigotimes$& $A_{1g}$ & $A_{1u}$ & $A_{2g}$ & $A_{2u}$ & $E_{u}$ & $E_{g}$\\
    \hline
    $A_{1g}$ & $A_{1g}$ & $A_{1u}$ & $A_{2g}$ & $A_{2u}$ & $E_{u}$ & $E_{g}$ \\
    $A_{1u}$ & - & $A_{1g}$ & $A_{2u}$ & $A_{2g}$ & $E_{g}$ & $E_{u}$ \\
    $A_{2g}$ & - & - & $A_{1g}$ & $A_{1u}$ & $E_{u}$ & $E_{g}$ \\
    $A_{2u}$ & - & - & - & $A_{1g}$ & $E_{g}$ & $E_{u}$ \\
    $E_{u}$ & - & - & - & - & $A_{1g} \oplus A_{2g} \oplus E_{g}$ & $A_{1u} \oplus A_{2u} \oplus E_{u}$ \\
    $E_{g}$ & - & - & - & - & - & $A_{1g}\oplus A_{2g} \oplus E_{g}$ \\
    \hline
\end{tabular}
\end{center}
\label{tab:directproduct}
\end{table}
When considering which of the coupling terms in the interaction potential were allowed, we used the direct product table of the point group $D_{3d}$ (Table \ref{tab:directproduct}). As the potential energy $U$ is scalar and invariant under point-group operations, it transforms as the totally symmetric representation $A_{1g}$. Therefore, the formal selection rule for coupling terms to exist, is that the product of their irreducible representations contains the $A_{1g}$ representation \cite{PhysRevLett.122.073901, PhysRevLett.118.054101}. Here, it is important to note that the THz electric field lies in the $xy$ plane and therefore it transforms as $E_u$. For example, the nonlinear coupling term described by the interaction potential $\propto Q_\mathrm{R} Q_\mathrm{IR}^2$ involving the Raman-active $E_g$ phonon and an IR-active $A_{2u}$ is not allowed, since $E_g \otimes A_{2u} \otimes A_{2u} = E_g $ does not contain the totally symmetric representation. Moreover, we note that the linear excitation (electric dipole interaction) of the IR-active $A_{2u}$ described by the potential $\propto Q_{\mathrm{IR}} E$ is not allowed for fields in the $x-y$ plane (in-plane fields transform as $E_u$), and therefore we do not consider phonons of this species in the main article. It is only allowed when the field is along the $z$-axis (in which case the field transforms as $A_{2u}$). This fact is also intuitive, as the IR-active $A_{2u}$ mode corresponds to out-of-plane vibrations \cite{https://doi.org/10.1002/jrs.6255, PhysRevB.102.104102} and can therefore only be excited by an out-of-plane field. 

Regarding the terms involving the Raman-active $E_g$ phonon and the IR-active $E_u$ phonon, all terms discussed in the main article are allowed by symmetry.

\newpage 

\section{Details of 2D THz spectroscopy}
\label{sec:2Dspectroscopy}
\begin{figure}[h!]
    \centering
    \includegraphics{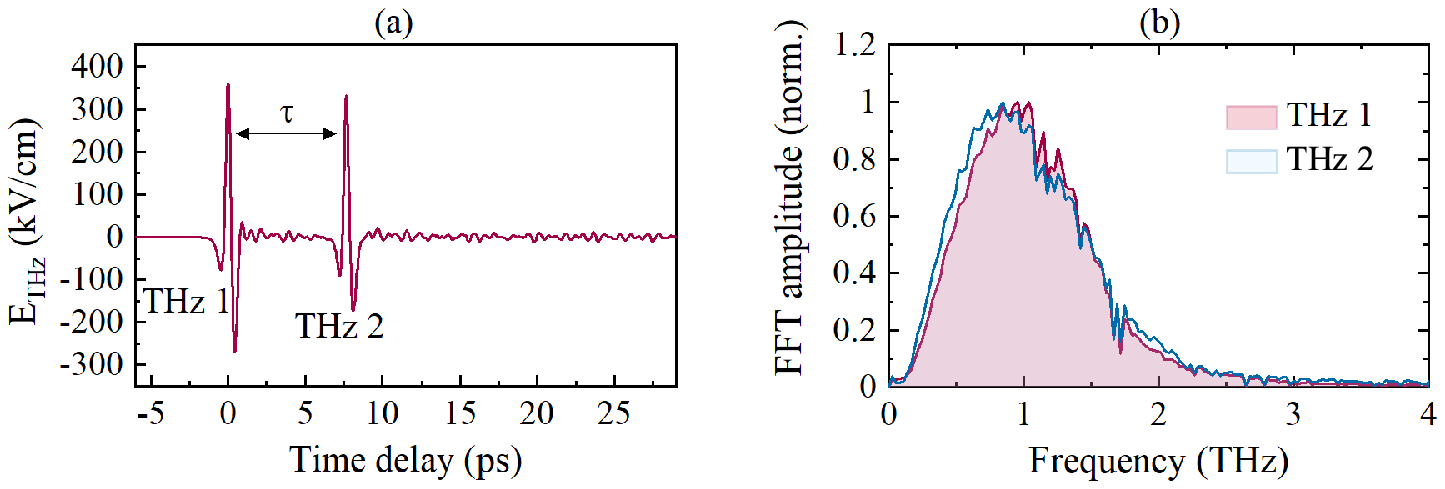}
    \caption{\textit{\small{(a) Calibrated waveforms of the two THz pump pulses used for 2D spectroscopy, measured by electro-optical sampling in GaP. (b) Associated normalized Fourier spectra of the two THz pulses. Although the two THz waveforms in the time domain are a bit different, their absolute Fourier spectra are nearly identical. }}}
    \label{fig:2dTHzcallib}
\end{figure}

In the double-pump experiment, we generated two THz pulses with peak electric fields of approximately $350$ and $330$~kV/cm (see Fig.~\ref{fig:2dTHzcallib}) in a single LiNbO\textsubscript{3} crystal from two separate optical pulses with a central wavelength of $800$~nm, whose mutual delay (the excitation time $\tau$) can be adjusted. We modulated the THz pump $1$ and pump $2$ at unequal repetition rates of $500$ and $250$~Hz, respectively, by mechanical choppers. Thereby, we could isolate the signal from pump $1$ ($\theta_1(t,\tau)$) and pump $2$ ($\theta_2(t))$ separately with a data-acquisition card. However, due to the non-linear nature of the generation process by optical rectification, the simultaneous presence of the two pulses inside the LiNbO\textsubscript{3} crystal ($\tau \approx 0$) results in a parasitic mutual interaction between the pulses, which affects the generation process. This causes anomalous results around the pump-pump overlap $\tau = 0$, as superposition is broken by the generation process itself. Therefore, during the analysis of the results (when taking the 2D FFT of the data) presented in the main article, we excluded a small region around the overlap of the two THz pulses $|\tau| < 1$~ps, in order to exclude any artificial signals. 
\newpage 

\section{Analytical solution to the equations of motion}
\label{sec:analytical}
The THz electric field can be modeled accurately by the Gaussian derivative function:
\begin{equation}
    \label{eq:Gaussderiv}
    E(t) = t_0 \frac{d}{dt} \mathcal{G}(t), \qquad \mathcal{G}(t) = E_0 \sin\omega_0 t \exp\left(-\frac{t^2 }{t_0^2}\right)
\end{equation}
where $\omega_0 t_0 \approx 1$ and $E_0$ is the peak electric field strength. This electric field is the direct driving force of the IR-active phonon modes described by the normal coordinate $Q_\eta$, whose equations of motion in absence of dissipation are of the form:
\begin{equation}
    \ddot{Q}_{\eta} + \omega_\eta^2Q_{\eta} = \frac{Z_\eta^*E(t)}{M_\eta}
\end{equation}
In order to solve this equation, we first substitute $\xi_\eta = \dot{Q}_\eta + i\omega_\eta Q_\eta$ and solve the resulting first order differential equation for $\xi_\eta$. Then we can obtain our solution $Q_\eta = \frac{1}{\omega_{\eta}}\Im{\xi_\eta}$, which gives:
\begin{eqnarray}
    \label{eq:solution_nonimpulsive}
    Q_\eta(t)\rvert_{t\gg t_0} &=& \frac{Z_\eta^*}{\omega_\eta M_\eta} \Im{\exp(i\omega_\eta t)\int_{-\infty}^ \infty E(t) \exp(-i\omega_\eta t) \, \mathrm{d}t }
\end{eqnarray}
after integration by parts:
\begin{eqnarray}
    Q_\eta(t)\rvert_{t\gg t_0} &=& \frac{Z_\eta^* E_0 t_0 }{ M_\eta}\Im{i \exp(i\omega_\eta t)\int_{-\infty}^ \infty \sin\omega_0 t  \exp(-t^2/t^2_0) \exp(-i\omega_\eta t) \, \mathrm{d}t} \\
    &=& \frac{Z_\eta^* E_0 t_0 }{ 2M_\eta}\Im{ \exp(i\omega_\eta t)\int_{-\infty}^ \infty \left[\cos(\omega_0 - \omega_\eta) t - \cos(\omega_0 + \omega_\eta) t\right] \exp(-t^2/t^2_0) \, \mathrm{d}t} \nonumber
\end{eqnarray}
where we used that thy symmetric integration over an odd function around zero is zero. After the substitution $t' \mapsto \sqrt{2}t/t_0$, we can apply the identity $\int_{-\infty}^\infty \cos(ax)\exp(-\tfrac{1}{2}x^2)\, \mathrm{d}x = \sqrt{2\pi} \exp(-\tfrac{1}{2}a^2)$ and obtain after rearranging the terms:
\begin{eqnarray}
 Q_\eta(t)\rvert_{t\gg t_0} &=& \frac{\sqrt{2\pi} Z_\eta^* E_0 t^2_0 }{ 2M_\eta} \left[\exp(-\tfrac{1}{4}(\omega_0 - \omega_\eta)^2t_0^2) - \exp(-\tfrac{1}{4}(\omega_0 + \omega_\eta)^2t_0^2)\right]\sin\omega_\eta t  \nonumber \\
 &=& \frac{Z_\eta^* E_0 t^2_0 }{ M_\eta} \Xi (\omega_\eta) \sin\omega_\eta t \label{IR-mode-solution}
\end{eqnarray}
where $\Xi(\omega) \equiv \sqrt{\pi}\exp(-\tfrac{1}{4} (1+t_0^2\omega^2))\sinh(\tfrac{1}{2}t_0\omega)$. For later purpose we define $K_\eta = \frac{ Z_\eta^*}{M_\eta} \Xi(\omega_\eta)$.
Now, we are interested in the nonlinear excitation of the $E_g$ Raman-active phonon through these IR-active modes. When an IR-active phonon with normal coordinate $Q_\eta$ is excited by the first THz pulse (electric field $E(t)$), the second THz pulse (electric field $E'(t,\tau)$) delayed at a time $\tau$ may excite the Raman-active phonon through nonlinear coupling term $\propto Q_\mathrm{R} Q_\mathrm{IR} E$. The equation of motion for the Raman-active phonon (with normal coordinate $Q_R$), taking into account this nonlinear coupling term, is given by:
\begin{equation}
    \label{Raman_eqofmotion}
    \ddot{Q}_\mathrm{R} + \omega_\mathrm{R}^2Q_\mathrm{R} = \frac{1}{M_\mathrm{R}} \sum_{\eta=1}^4 b_\eta Q_{\mathrm{IR}\eta}(t)E'(t, \tau)
\end{equation}
where the $b_\eta$ coupling constants. The electric field of the second THz pulse with a peak amplitude of $E_0'$ is similar to the first THz pulse, but delayed at a time $\tau>0$:
\begin{equation}
    \frac{E'(t, \tau)}{E'_0} = \frac{E(t-\tau)}{E_0}
    \label{THz2}
\end{equation}
This second pulse will efficiently excite the Raman-active mode via the already excited IR-active mode of the first pulse. Note that for simplicity, we ignore the occurrence of this process within a single THz pulse. The solution for $t\gg \tau$ of the Raman-active mode normal coordinate in Eq.~\ref{Raman_eqofmotion} is given by:
\begin{align}
    Q_\mathrm{R}(t)\rvert_{t\gg \tau} &= \frac{1}{\omega_\mathrm{R} M_\mathrm{R}}\sum_{\eta=1}^4b_\eta\Im{e^{i\omega_\mathrm{R} t}\int_{-\infty}^ \infty Q_{\mathrm{IR}\eta}(t)E'(t) e^{-i\omega_\mathrm{R} t} \, \mathrm{d}t } \\
    &= \frac{ E_0E'_0t_0^3}{\omega_\mathrm{R} M_\mathrm{R}}\sum_{\eta=1}^4 b_\eta K_\eta   \Im{e^{i\omega_\mathrm{R} t}\int_{-\infty}^\infty \sin\omega_\mu t \frac{\mathrm{d}}{\mathrm{d}t}\left[\sin\omega_0(t-\tau) \exp(-\frac{(t-\tau)^2}{t_0^2})\right]e^{-i\omega_\mathrm{R} t}\, \mathrm{d}t} \nonumber \\
    &= \frac{ E_0E'_0t_0^3}{2\omega_\mathrm{R} M_\mathrm{R}}\sum_{\eta=1}^4 b_\eta K_\eta  \Bigg[  \Im{-(\omega_\mathrm{R}+\omega_\eta)e^{i\omega_\mathrm{R} t}e^{-i(\omega_\mathrm{R} + \omega_{\eta})\tau}  \int_{-\infty}^\infty \sin\omega_0t e^{-\frac{t^2}{t_0^2}} e^{-i(\omega_\mathrm{R} + \omega_\eta)t} \, \mathrm{d}t}  \nonumber\\ 
    & \qquad \qquad \qquad \qquad    + \Im{(\omega_\mathrm{R}-\omega_\eta)e^{i\omega_\mathrm{R} t} e^{-i(\omega_\mathrm{R} - \omega_{\eta})\tau} \int_{-\infty}^\infty \sin\omega_0t e^{-\frac{t^2}{t_0^2}} e^{-i(\omega_\mathrm{R} - \omega_\eta)t} \, \mathrm{d}t } \Bigg] \nonumber \\
    &= E_0E'_0t_0^4\sum_{\eta=1}^4 b_\eta K_\eta  \Bigg(L^+_{R\eta} \cos(\omega_\mathrm{R} t - (\omega_\mathrm{R} + \omega_\eta) \tau)  - L^-_{R\eta} \cos(\omega_\mathrm{R} t - (\omega_\mathrm{R} - \omega_\eta) \tau)\Bigg) \label{solutionRaman1}
\end{align}
where $L^\pm_{R\eta} = \frac{\omega_\mathrm{R} \pm \omega_\eta}{\omega_\mathrm{R} M_\mathrm{R}}\Xi(\omega_R \pm \omega_\eta) $. 
Therefore, we will observe peaks in the 2D FFT spectrum at the complementing excitation frequencies $(\omega_{\mathrm{det}}, \omega_{\mathrm{ex}}) = (\omega_\mathrm{R}, \omega_\mathrm{R} \pm \omega_\eta)$.

To explain the peak seen at the excitation frequency of the IR-active phonon frequency, i.e. at $(\omega_{\mathrm{det}}, \omega_{\mathrm{ex}}) =(\omega_\mathrm{R}, \omega_\eta)$, we have to look at the region where moving THz pulse $E'(t)$ arrives \textit{before} the ''first'' THz pulse $E(t)$ (when $\tau < 0$). Therefore, we first need to solve the excitation of the IR-active mode with pulse $E'(t-\tau)$ with $\tau<0$ as in Eq.~\eqref{THz2}, which gives:
\begin{equation}
    Q_\eta(t)\rvert_{t> \tau, t<0} = \frac{Z_\eta^* E'_0 t^2_0 }{ M_\eta} \Xi (\omega_\eta) \sin\omega_\eta (t-\tau). 
    \label{eq:IRphonon2}
\end{equation}
Afterward, we again solve Eq.~\eqref{Raman_eqofmotion} with $E(t)$ as in Eq.~\eqref{eq:Gaussderiv}, in the presence of an IR-active phonon described by Eq.~\ref{eq:IRphonon2}. The solution is given by substituting $t \mapsto t+\tau$ in the solution of Eq.~\eqref{solutionRaman1}:
\begin{equation}
    Q_\mathrm{R}(t)\rvert_{t> \tau, t>0} = E_0E'_0t_0^4\sum_{\eta=1}^4 b_\eta K_\eta  \Bigg(L^+_{R\eta} \cos(\omega_\mathrm{R} t - \omega_\eta \tau)  - L^-_{R\eta} \cos(\omega_\mathrm{R} t + \omega_\eta \tau)\Bigg).
\end{equation}
It means that depending on whether the moving pulse comes first or second, one can distinguish between the real and complementing phonon peak. If one takes the 2D FFT where the moving pulse comes first $\tau < 0$, a real phonon peak will appear in the spectrum. And if the Fourier transform is taken over the region where the moving pulse comes after THz pulse at $t=0$ (i.e. $\tau > 0$), the complementary phonon peak $\omega_\mathrm{R} - \omega_\eta$ will be present in the spectrum. However, due to the internal reflection in our experimental data, a moving pulse always comes before and after and we cannot test this hypothesis experimentally. But we can check our hypothesis with the numerical simulations. This is done in Section~\ref{sec:Distinguishing-peaks}, and it confirms our expectation.

\newpage

\section{Numerical simulations of phonon dynamics}
\begin{figure}[h!]
    \centering
    \includegraphics{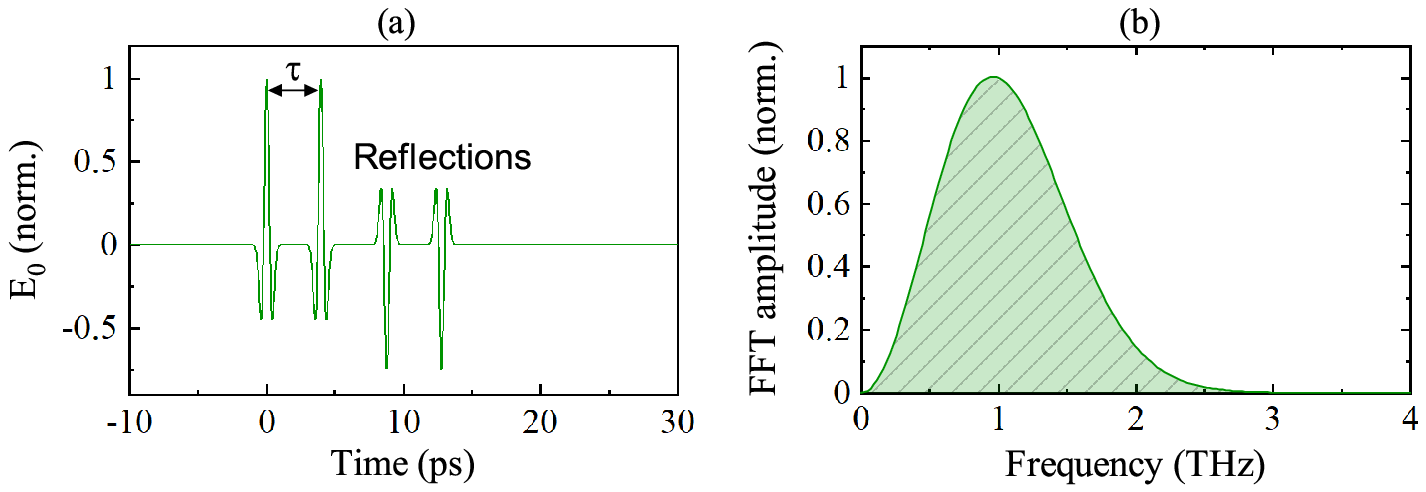}
    \caption{\textit{\small{(a) Simulated THz pulses by the Gaussian derivative function, including the reflection to resemble the experimental situation. (b) The simulated THz spectrum, which is highly similar to the actual spectrum shown in Fig.~\ref{fig:2dTHzcallib}(b).}}}
    \label{fig:2Dsimulated-waveforms}
\end{figure}

In the simulations, we model the two THz electric field pulse shapes by the Gaussian derivative function~\eqref{eq:Gaussderiv}, using the parameters $t_0 = 0.36$~ps, $\omega_0 = 1/t_0$ and the electric field $E_0$ set to unity. The simulated waveforms and their associated spectra are shown in Fig.~\ref{fig:2Dsimulated-waveforms}.

The two coupled equations of motion that were simulated are:
\begin{eqnarray}
    \ddot Q_{\mathrm{IR}} + 2\zeta_{\mathrm{IR}}\omega_{\mathrm{IR}} \dot Q_{\mathrm{IR}} + \omega_{\mathrm{IR}}^2 Q_{\mathrm{IR}}  &=& \frac{Z^*}{M_{\mathrm{IR}}}E + \frac{1}{M_\mathrm{IR}}b Q_\mathrm{R}E+ \frac{2}{M_\mathrm{IR}}cQ_\mathrm{R}Q_\mathrm{IR}\label{IRRS1} \\ 
    \ddot Q_\mathrm{R} + 2\zeta_{\mathrm{R}}\omega_{\mathrm{R}} \dot Q_{\mathrm{R}} + \omega_\mathrm{R}^2 Q_\mathrm{R}  &=& \frac{\delta_\mathrm{R}}{M_\mathrm{R}} E^2 + \frac{1}{M_\mathrm{R}} b Q_{\mathrm{IR}}E + \frac{1}{M_\mathrm{R}}Q_{\mathrm{IR}}^2 \label{IRRS2}
\end{eqnarray}
where we took a Raman-active phonon with a frequency of $\tfrac{1}{2\pi} \omega_\mathrm{R} = 3.14$~THz and an IR-active phonon with a frequency of $\tfrac{1}{2\pi} \omega_{\mathrm{IR}} = 1.47$~THz. The damping ratios were set at $\zeta_{\mathrm{R}} = 0.005$ and $\zeta_{\mathrm{IR}} = 0.015$ and the effective masses were put to unity $M_{\mathrm{IR}} = M_{\mathrm{R}} = 1$ as well as the effective charge $Z^* = 1$. In the following sections, we display the numerical solutions for each of the three nonlinear coupling terms in Eq. \eqref{IRRS2}, separately. 
In all cases, the relevant parameter ($b, c, \delta_\mathrm{R})$ was set to $1$.

The equations were solved numerically using the ordinary differential equation integrator \textit{scipy.integrate.}\textit{odeint} in Python. Just as in the experiment, we can take the signal with only one of the pulses present ($Q_\mathrm{R, 1}$ and $Q_\mathrm{R,2}$) or both pulses present ($Q_{\mathrm{R},12}$) and calculate the by definition nonlinear part:
\begin{equation}
     Q_{\mathrm{R, NL}}(t, \tau) = Q_{\mathrm{R},12}(t,\tau) - Q_{\mathrm{R},1}(t, \tau) - Q_{\mathrm{R},2}(t, \tau), \label{NLsim}
\end{equation}

\subsection{THz-mediated coupling between the modes $\propto Q_{\mathrm{R}}Q_{\mathrm{IR}} E$}
\label{sec:IRRS} 

\begin{figure}[h!]
    \centering
    \includegraphics{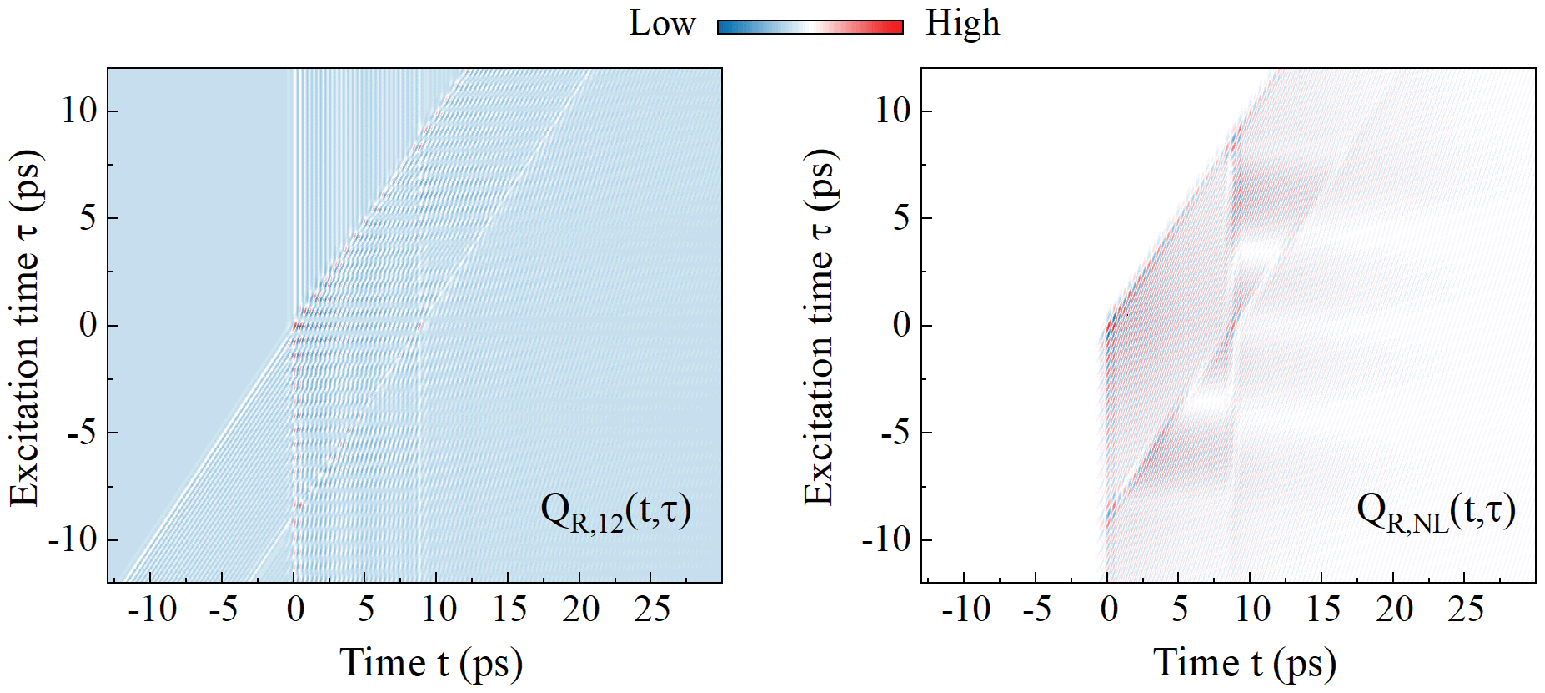}
    \caption{\textit{\small{Simulated results of $Q_\mathrm{R}$ for the driving force $\propto Q_{\mathrm{IR}} E$ $(b=1)$ in time domain, of both the total signal $Q_{\mathrm{R}, 12}$ and the non-linear signal $Q_{\mathrm{R, NL}}$. }}}
    \label{fig:IRRS_TD}
\end{figure}

\begin{figure}[h!]
    \centering
    \includegraphics{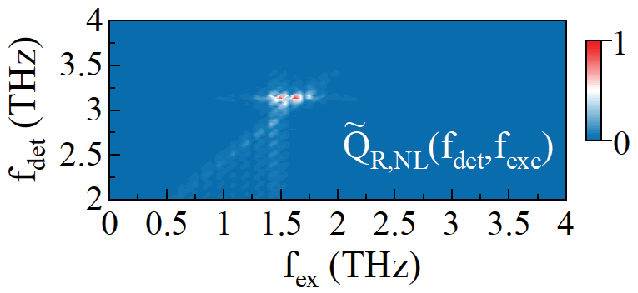}
    \caption{\textit{\small{Normalized 2D FFT associated to Fig.~\ref{fig:IRRS_TD}.}}}
    \label{fig:IRRS_FD}
\end{figure}

\newpage 
\subsection{Photonic impact $\propto E^2$}
\begin{figure}[h!]
    \centering
    \includegraphics[scale = 0.9]{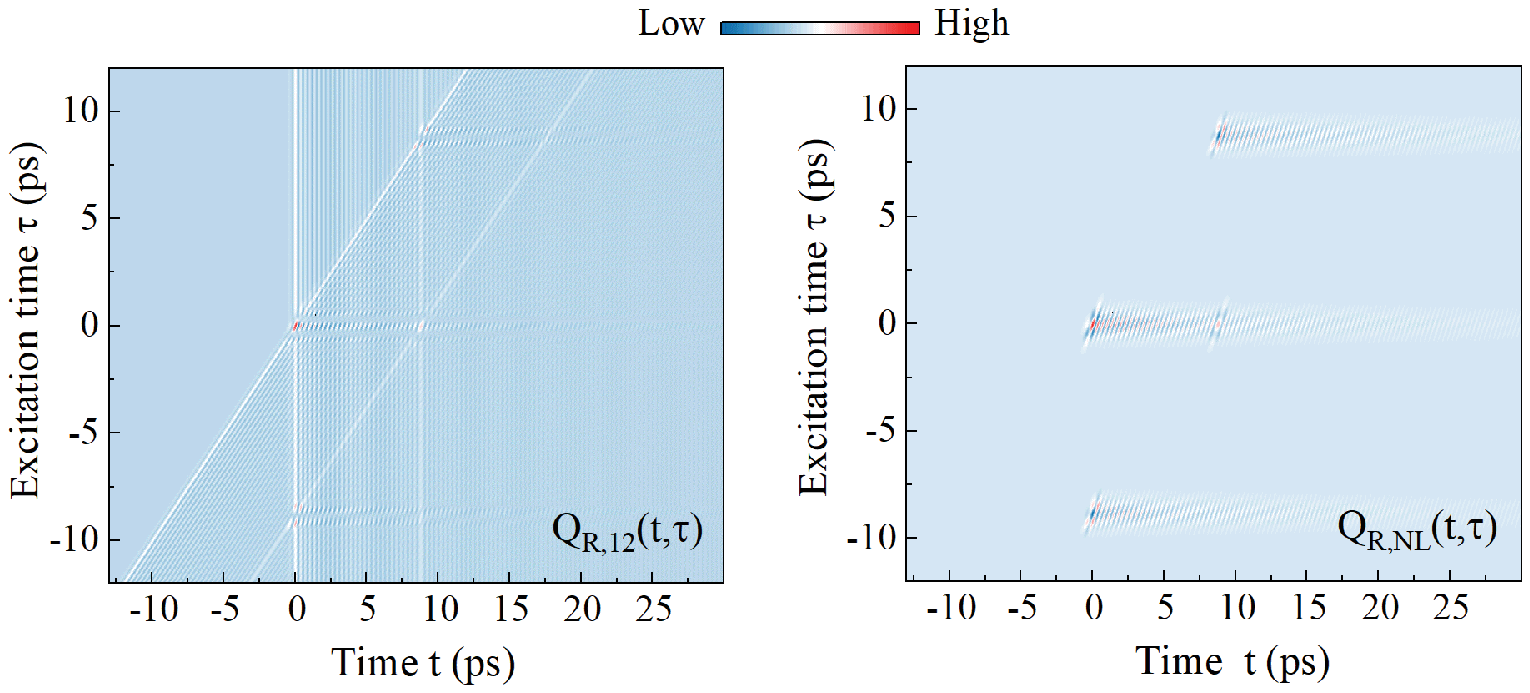}
    \caption{\textit{\small{Simulated results when taking only photonic impact $\propto E^2$ into account $(\delta_\mathrm{R}=1)$. Both the total signal $Q_{\mathrm{R}, 12}$ and the non-linear signal $Q_{\mathrm{R, NL}}$ are depicted.}}}
    \label{fig:2pa}
\end{figure}

\begin{figure}[h!]
    \centering
    \includegraphics{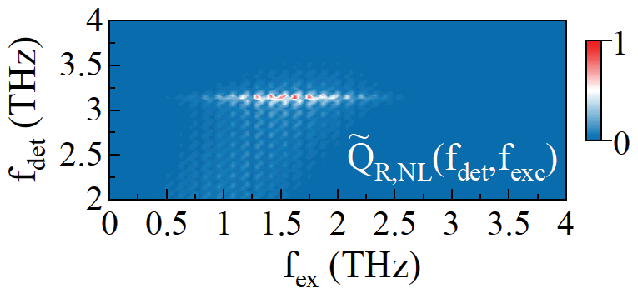}
    \caption{\textit{\small{Normalized 2D FFT associated to Fig.~\ref{fig:2pa}.}}}
    \label{fig:2pa_FD}
\end{figure}

The result for the photonic impact term is shown in Fig.~\ref{fig:2pa}. Regarding the nonlinear part, there is only a signal around the pulse overlap $\tau = 0$. This is in stark contrast with the experimental result, as the experiment showed that the long-lived IR-active phonons can mediate an interaction between two subsequent THz pulses over pump-pump separation times $\tau$ much longer than the pulse duration itself. Additionally, when taking the 2D FFT over the entire time domain, we obtain the result presented in Fig.~\ref{fig:2pa_FD}. It shows a broad line in the excitation spectrum, which clearly does not produce the experimental result. Moreover, in the analysis of the experimental data, we exclude the region around the pulse overlap (see Section~\ref{sec:2Dspectroscopy}), meaning that we removed most of 
the photonic impact contributions to the signal (if present). 
Therefore, we exclude photonic impact as the dominant mechanism describing our data.

\newpage

\subsection{Anharmonic lattice dynamics $\propto Q_{\mathrm{R}} Q_{\mathrm{IR}}^2$}
\begin{figure}[h!]
    \centering
    \includegraphics[scale = 0.9]{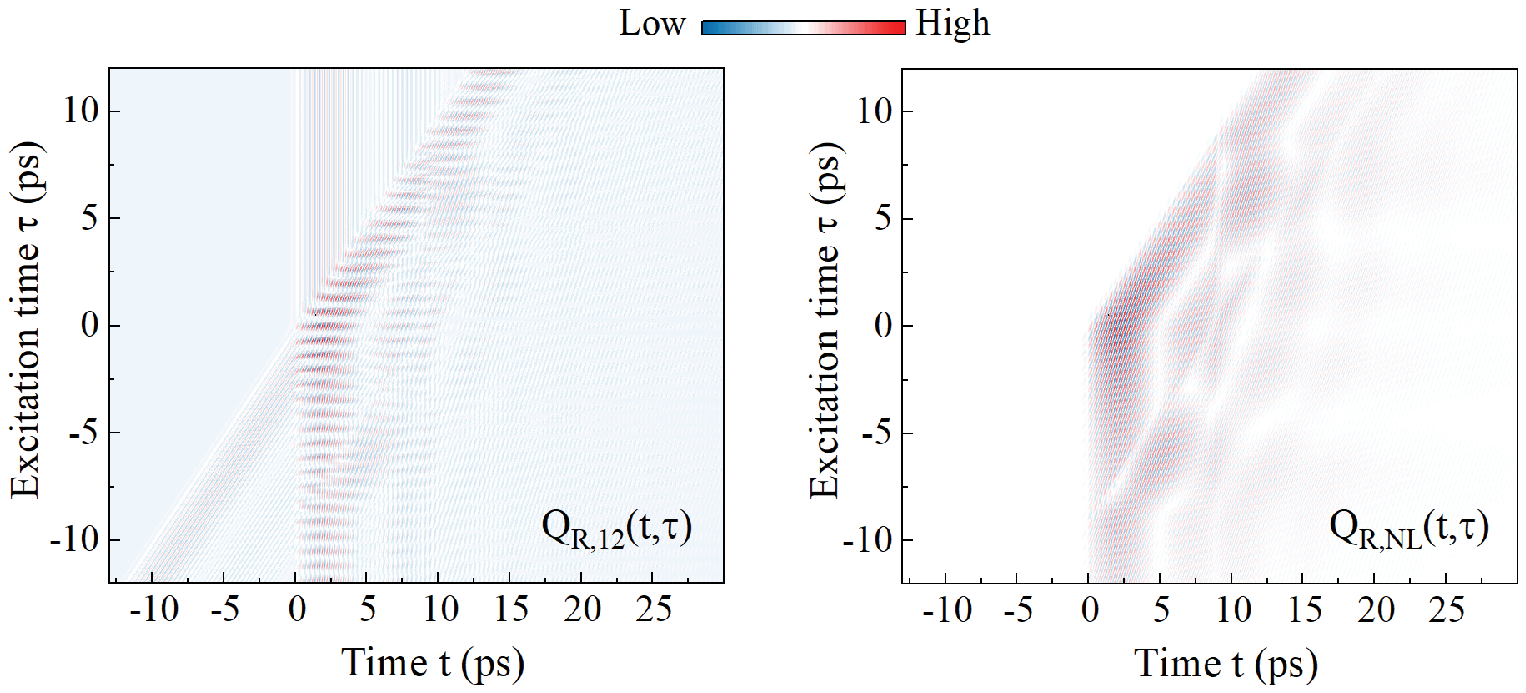}
    \caption{\textit{\small{Simulation of $Q_\mathrm{R, 12}(t, \tau)$ and $Q_\mathrm{R, NL}(t,\tau)$, taking only the nonlinear coupling term  $\propto Q_{\mathrm{R}} Q_{\mathrm{IR}}^2$ into account ($c=1$).}}}
    \label{fig:iRS}
\end{figure}

\begin{figure}[h!]
    \centering
    \includegraphics{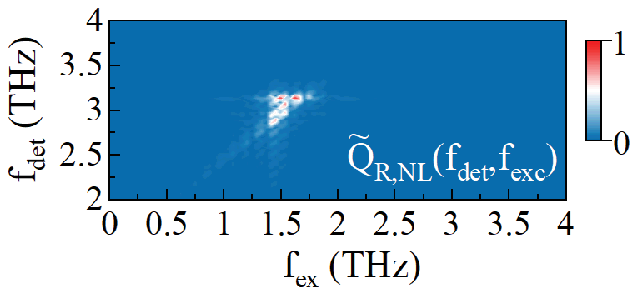}
    \caption{\textit{\small{Normalized 2D FFT associated to Fig.~\ref{fig:iRS}.}}}
    \label{fig:IRS_FD}
\end{figure}

The simulation result for the driving term $\propto Q_{\mathrm{IR}}^2$ for the Raman-active phonon is depicted in Fig.~\ref{fig:iRS}. In the corresponding 2D FFT shown in Fig.~\ref{fig:IRS_FD}, one can also see the two peaks, similar to the experimental result. As the driving term $\propto Q_{E_u}^2$ is not resonant with the $E_g$ mode frequency, i.e. $2f_{E_u} \neq f_{E_g}$, we can argue it effectively reduces to the THz-mediated phonon-phonon coupling mechanism as discussed in Section~\ref{sec:IRRS}. Namely, during the pulse overlap, the $Q_{E_u}$ normal coordinate undergoes a forced oscillation in which essentially $Q_{E_u}(t) \propto E_{\mathrm{THz}}(t)$. Therefore, while the THz pulse is present, one can effectively replace one $Q_{E_u}$ by $E$ in the anharmonic lattice term $Q_{E_u}^2$ to obtain the driving term $Q_{E_u} E$, which does obey the resonant condition due to the broad spectrum of the THz pulse. Hence, the same analytical analysis from Sec.~\ref{sec:analytical} holds (where we only considered the interaction potential $Q_\mathrm{R}Q_{\mathrm{IR}}E$. However, in addition to the two peaks, we see that the anharmonic lattice term in Fig.~\ref{fig:IRS_FD} also produces a large spectral component away from the detection frequency of the $E_g$ phonon, something which is not seen in the experiment. This is because the presence of an IR-phonon constantly forces the oscillation of the Raman-active mode, and exists also when the THz electric field is not present. Therefore, although this driving term also reproduces the two peaks, we exclude it as the dominant mechanism. 

\newpage

\subsection{Distinguishing phonon and complementing peaks}
\label{sec:Distinguishing-peaks}
\begin{figure}[h!]
    \centering
    \includegraphics{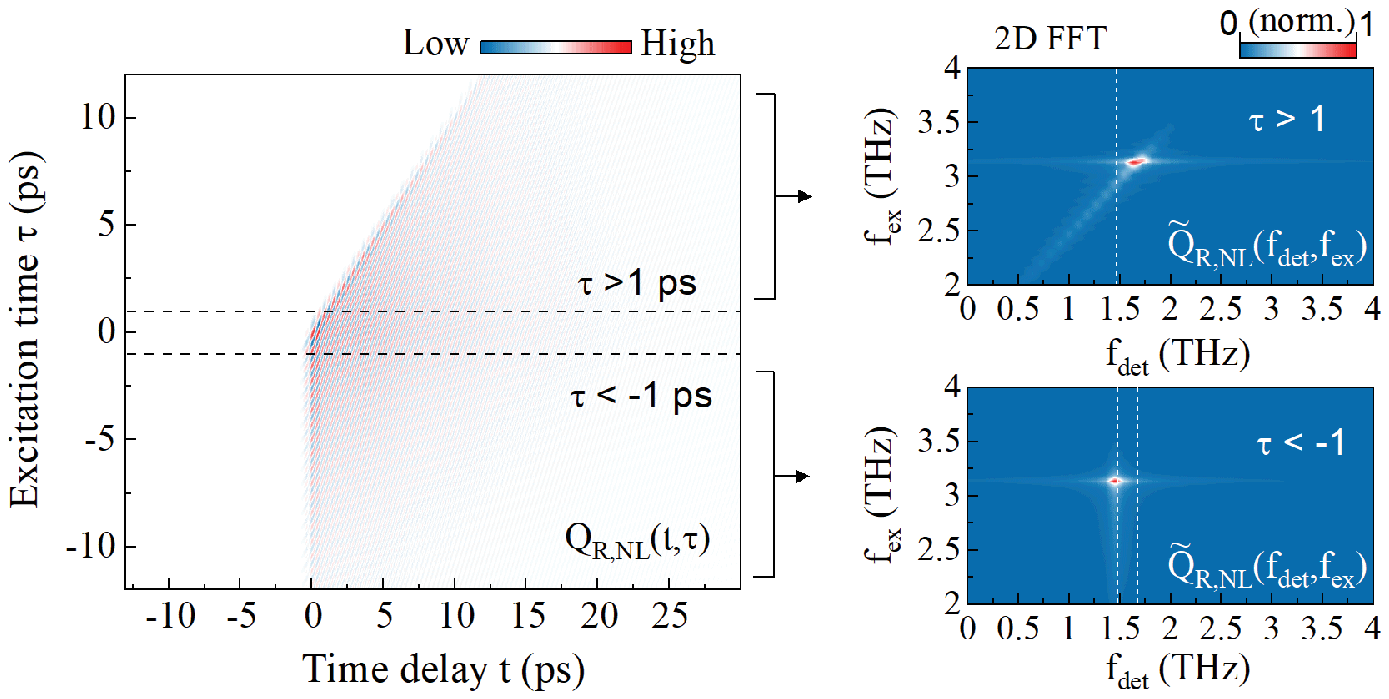}
    \caption{\textit{\small{Simulated results as in Fig.~\ref{fig:IRRS_TD} (interaction potential $\propto Q_{\mathrm{R}}Q_{\mathrm{IR}}E$), but without the internal THz reflection. Taking the 2D FFT in the region where the moving pulse arrives later $\tau>1$ shows the complementary phonon peak in the frequency domain. However, when making the 2D FFT transformation in the region where the moving pulse arrives first ($\tau<-1$), we obtain the actual phonon frequency in the spectrum.}}}
    \label{fig:IRRSbeforeafter}
\end{figure}

By deriving the analytical solutions of the equations of motion (Section~\ref{sec:analytical}), we hypothesized the possibility to distinguish the actual phonon peak from the complementary ones when considering the THz-mediated mode coupling term ($\propto Q_{\mathrm{R}}Q_{\mathrm{IR}}E$), by taking the FFT over the negative and positive $\tau$, respectively. However, this effect is hampered experimentally by internal reflections. Therefore, we repeated the numerical simulation as in Sec.~\ref{sec:IRRS}. but without the simulated internal reflections. The results by taking the FFT over negative and positive $\tau$ confirm our hypothesis, as can be seen in Fig. \ref{fig:IRRSbeforeafter}.

\newpage

\bibliography{references}